# Low-Force Elastocaloric Refrigeration via Bending

**AUTHORS:** *Darin J. Sharar[1*], Joshua Radice[2], Ronald Warzoha[2], Brendan Hanrahan[1], Andrew Smith[2]*

**AFFILIATIONS**

[1]U.S. Army Research Laboratory, Adelphi, MD 20783, United States

[2]U.S. Naval Academy, Annapolis, MD 21402, United States

**CORRESPONDING AUTHOR:** *Correspondence and requests for materials should be addressed to D. Sharar (darin.j.sharar.civ@mail.mil)

**ABSTRACT**

Elastocaloric cooling has been identified as a promising alternative to high global warming potential vapor compression cooling. Two key bottlenecks to adoption are the need for bulky/expensive actuators to provide sufficient uniaxial stress and inadequate elastocaloric material fatigue life. This paper defines the physics that govern performance of axisymmetric flexural bending for use as an emerging low-force and low-fatigue elastocaloric heating and cooling mechanism and further demonstrates a continuous rotary-driven cooling prototype using polycrcrystalline $Ni_{50.7}Ti_{48.9}$. Elastocaloric material performance is determined using infrared thermography during uniaxial-tension and four-point bending thermomechanical testing. A systematic study reveals the effects of strain rate (from 0.001 to 0.025 $s^{-1}$), maximum strain (from 2 to 8%), and strain mode on the temperature evolution, mechanical response, and coefficient of performance. Four-point bending experiments demonstrate a temperature reduction up to 11.3°C, material coefficients of performance between 2.31 and 21.71, and a 6.09- to 7.75-fold reduction in required actuation force compared to uniaxial tension. The absence of Lüders bands and reduced mechanical dissipation during flexure represent reduced microstructure degradation and improved fatigue life. The rotary-based elastocaloric cooling prototype is shown to provide similar thermomechanical performance with the added benefit of discrete hot and cold zones, continuous cooling, inexpensive rotary actuation, and scalability, which represents a significant advancement for compact, long lifetime, and inexpensive elastocaloric cooling.

**MAIN TEXT**

Important climate change legislation has been proposed in the United States, as well as Canada, Mexico, and the European Union, to phase out high global warming potential (GWP) Hydrofluorocarbon (HFC) refrigerants used in vapor-compression (VC) cooling [1]. Solid-state, zero-GWP alternatives including magnetocalorics, electrocalorics, and elastocalorics (eCs) are being actively pursued. Elastocalorics, which exchange mechanical and thermal energy via structural entropy changes, offer a promising approach with theoretical and observed COPs greater than 10 [2], large endothermic temperature changes up to 58K [3], operation up to 83% Carnot efficiency [4], and high volumetric energy density [5].

Most elastocaloric studies, both experimental and theoretical, have focused on shape memory alloy (SMA) development/testing [6] [7] [8] [9] and thermodynamic cooling cycles [10] [11] [12] [13]. Only a handful of researchers have made efforts to develop eC cooling devices [14] [15] [16] [17] [18] [19] [12] [20] [21]. The majority of these demonstrators have relied on straining the material uniaxially [14] [15], while others have accomplished uniaxial strain by either stretching thin films across convex surfaces [19] [21] or using composite structures that shift the neutral axis to illicit pure tension or compression [20]. In addition to requiring large force [22], such uniaxial approaches have generally been limited to $10^2$-$10^5$ cycles before catastrophic failure as a result of crack formation and propagation in the material due to repeated strain-induced phase nucleation and annihilation near Lüders bands [23] or engineering failure due to material delamination, for example [20].

To date, there are no reported systematic studies of axisymmetric bending-mode elastocaloric cooling whereby a cross section of material experiences both tension and compression, nor are there proposed system architectures available in the literature. Here we build on a preliminary work by the authors [24] to first compare the physics of bending-mode eC cooling to uniaxial tension and subsequently evaluate the thermomechanical performance characteristics of a novel elastocaloric cooling architecture that relies on continuously driving/ feeding a polycrystalline wire around a fixed circular surface to induce continuous bending-mode cooling. Critically, we find that axisymmetric bending enables a five-fold reduction in the required force for an equivalent COP and temperature lift, which directly corresponds to a reduction in the size, weight, and power input required for elastocaloric cooling systems. Furthermore, insights revealed by Lüders band formation and mechanical dissipation effects suggest bending may improve fatigue life compared to more-traditional uniaxial actuation approaches. In aggregate, these results represent an approach to mitigating key bottlenecks that have hindered past eC designs.

Superelastic $Ni_{50.7}Ti_{48.9}$ wire (henceforth referred to as NiTi) with an Austenite finish temperature between 10 and 18°C was purchased from Fort Wayne Metals. The material had less than 0.25 wt. % of trace elements such as Carbon, Hydrogen, Nitrogen, Oxygen, Cobalt, Copper, Chromium, Iron, and Niobium [25]. The samples were received and tested in polished straight-annealed form. As shown in Figure 1a-b, the uniaxial tension and four-point bending studies were performed in an ADMET single-column testing system at strain rates of 0.001, 0.0025, 0.01, and 0.025 $s^{-1}$ and strains of 2, 4, 6, and 8%, yielding 16 unique testing conditions for each case. The tensile tester was operated in displacement mode to tightly control the strain and the strain rate. After mechanically loading the sample, and before unloading, the strain/displacement was kept constant for 60 s to allow the exothermic latent heat associated with the Austenite→Martensite phase transformation to dissipate. A FLIR SC8300 infrared camera with a sensitivity of 0.025K was used to continuously measure the temperature evolution during the loading and unloading cycles. Materials were coated with graphite lubricant to provide constant emissivity (~0.93) across the temperature range of interest [26].

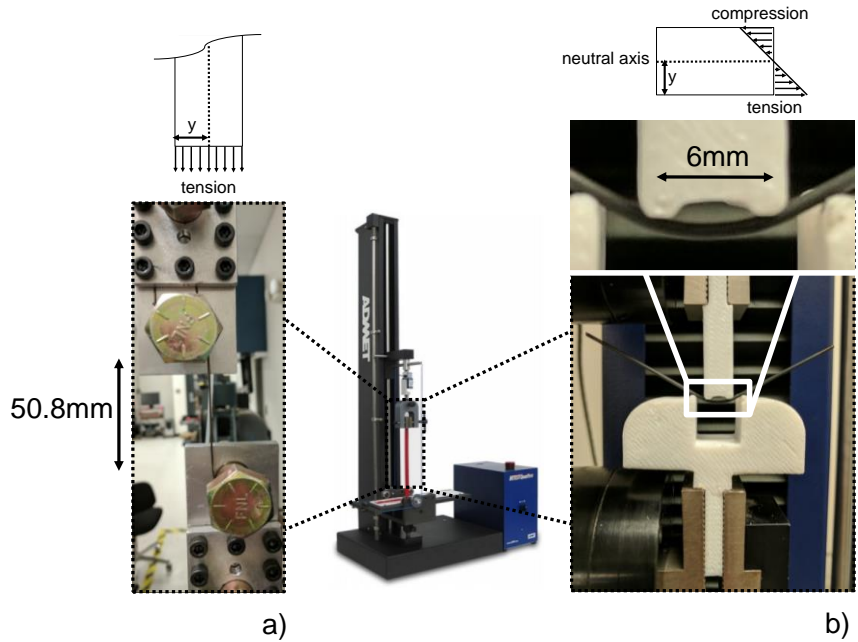

Figure 1: Photograph of the custom test fixtures made to interface with the ADMET single-column tester to allow a) uniaxial tension testing and b) four-point bending testing

The uniaxial tension fixture (Figure 1a) consisted of a caul plate and loop design, necessary to provide sufficient friction between the fixture and NiTi material (~50.8 mm active material) and prevent slipping during loading. Compression tests were not performed due to known buckling concerns [8]. Figure 1b is a photograph of the 3-D printed four-point bending test fixture; one clear benefit of bending is revealed here, whereby large clamp strength and buckling are avoided and tooling can be fabricated out of low-cost and low-weight materials. In the four-point bending configuration, the maximum flexural stress and strain is spread over the section of the NiTi sample between the top loading points shown in Figure 1b. This provides ~6 mm of NiTi material that is loaded at the same maximum stress and strain. Additionally, the majority of the actively strained area is not in contact with the polycarbonate anvil (0.19-0.23 $Wm^{-1}K^{-1}$ [27]), so less thermal interaction between the fixture and sample is expected, thus providing a more-adiabatic condition.

Controlling engineering strain ($\varepsilon$) during uniaxial testing follows the relation, $\varepsilon=\Delta l/l$, provided a known sample length (l). In our experiments, an optical method was required to determine the necessary crosshead displacement to yield the desired bending strain. To accomplish this, the sample was mechanically loaded until the observed curvature matched the contour of a circle with a known radius using the expression, $\varepsilon = y/R$, where $y$ is the distance from the neutral axis (this is the radius of the sample in the case of maximum strain), and $R$ is the radius of curvature. Strain gauges were considered but were ultimately not implemented because of known issues with bonding to NiTi and potential to locally impede the martensitic transformation [20] [28].

Figure 2a-d summarizes the mechanical force vs. strain results for the uniaxial tension and four-point bending tests. As shown, the hysteresis curves changed with both maximum strain and strain rate. At the lowest strain tested (2%) and the full range of strain rates, superelasticity was not observed in either the tension or bending cases. This is consistent with results from the literature [20] [4] that show higher tensile strains are required to promote large scale stress-induced martensitic transformations. At increased strains of 4-8%, both bending and uniaxial tension results display the characteristic superelastic response and a force plateau as the strain increases. As the strain approaches 8%, the force begins to further increase, indicating recoverable elastic deformation of the martensite phase.

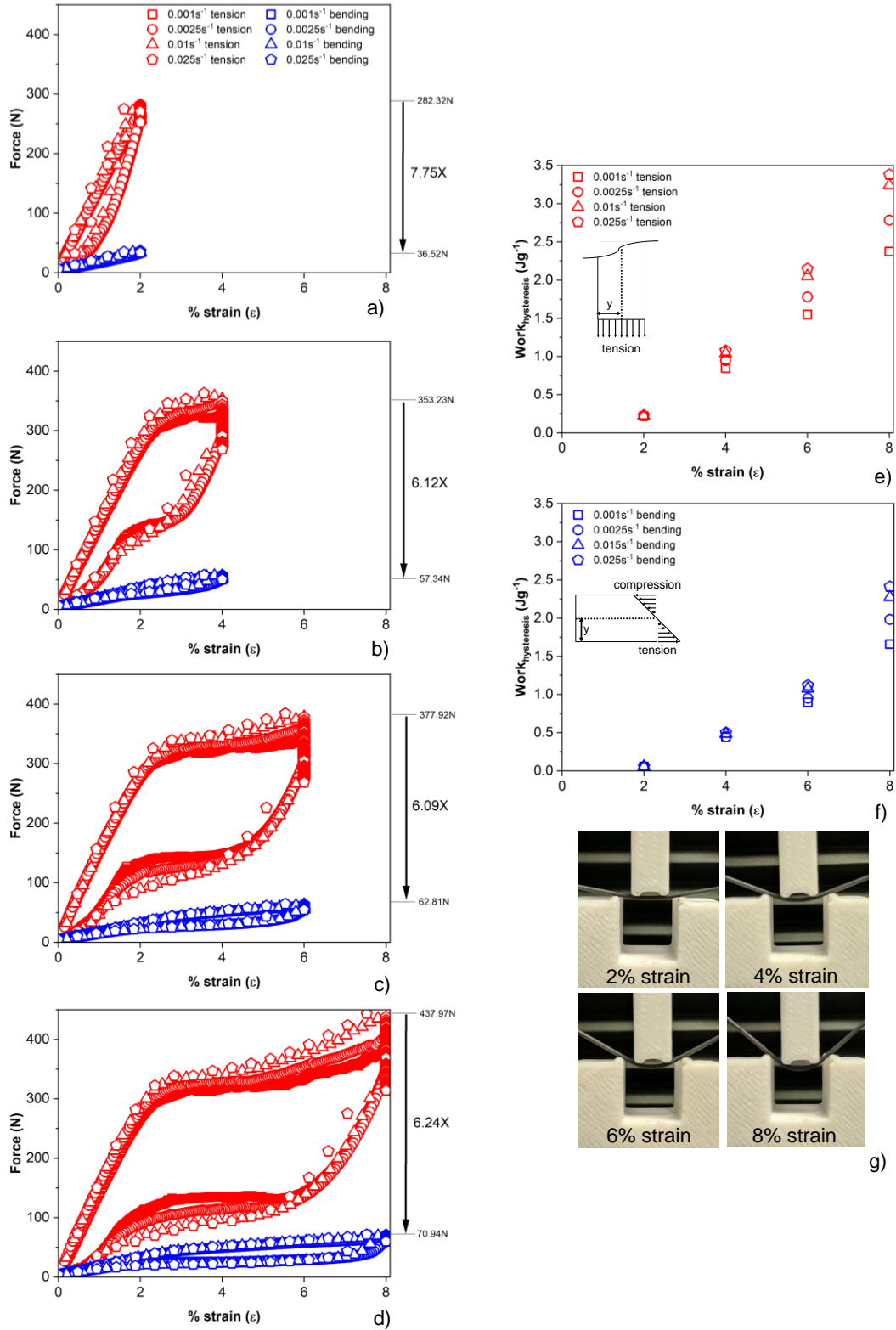

Figure 2: Mechanical testing results for all uniaxial and bending test cases; a)-d) are force vs % strain results for maximum strains of 2%-8%; e)-f) are dissipated mechanical energy (hysteresis work) for uniaxial tension and bending, respectively; g) is photographs of bending at strains of 2%-8%.

Generally, the maximum strain and energy required for an unloading-loading cycle (represented by the area inside the hysteresis loop) increased with increasing strain and strain rate. These results, normalized by the mass of active material, are summarized in Figure 2e-f for uniaxial tension and bending, respectively. Most notably, the maximum required force to obtain strains of 2, 4, 6, and 8% was 6.24-7.75 times smaller in the four-point bending tests compared to their uniaxial tension counterparts. Increased linear actuation distance for bending slightly reduced the resultant mass-specific energy requirements (Figure 2e-f) to 1.40-4.07 times lower than uniaxial tension.

As described in the literature [29] [21], the dependence of hysteresis on strain rate is due to self-heating and -cooling during loading and unloading, respectively, which results in a change in the critical stress values for phase transformation. This understanding is derived from the Clausius-Clapeyron relation:

$$\frac{dP}{dT} = \Delta L / (\varepsilon_0 T_C^0) \tag{1}$$

where $\Delta L$ is the latent heat, and $\varepsilon_0$ is the specific volume change during transformation. This relationship shows that the further away from the transition temperature, $T_C^0$, the greater the pressure dP required to induce the phase transformation. This is confirmed by Figure 3e-g which shows a representative temperature vs time response for the maximum (0.025 s$^{-1}$) and minimum (0.001 s$^{-1}$) strain rates tested for a strain of 6%. Infrared images corresponding to the highest and lowest temperatures for a strain rate of 0.025 s$^{-1}$ are superimposed on Figure 3a-b. In general, lower strain rates resulted in a more gradual phase transformation, a non-adiabatic response, and reduced self-heating and self-cooling. Alternatively, the highest strain rate of 0.025s$^{-1}$, approaching the adiabatic limit of 0.1 s$^{-1}$ reported elsewhere [21], resulted in large self-heating and self-cooling.

It's interesting to note the presence of Lüders bands in Figure 3a during uniaxial tension but not during bending in Figure 3b.

Figure 3d-e summarizes the endothermic and exothermic temperature responses during uniaxial tension (3d) and bending (3e). In both cases, the degree of self-heating and -cooling increased with increasing strain and strain rate as more of the material is transformed in a more-adiabatic condition. Thus, at a rate of 0.025 s$^{-1}$ and 8% max strain, the tension results show a peak exothermic and endothermic temperature response of 31.7 °C and -16.4 °C; based on the relation $\Delta L = C_P \Delta T$, with a measured specific heat ($C_P$) of 469 Jkg$^{-1}$K$^{-1}$ [30], these temperature changes corresponds to exo- and endothermic latent heats of 7.54 and 14.58 Jg$^{-1}$, respectively. We attribute the growing difference in endothermic and exothermic temperature and latent heat at higher strain rates to slip and other dissipating mechanisms commonplace to NiTi [31]. Furthermore, the temperature response increased linearly from 2 to 6%, but began to level off at 8%; presumably this is due to near-complete transformation of the material around 6% axial strain, as supported by the observed elastic deformation of the martensite phase (Figure 2d). Similarly, peak exothermic and endothermic temperature responses of 13.4 and -11.3°C occur at the maximum strain and strain rate during bending; the reported latent heat values are 6.16 and 5.20 Jg$^{-1}$, respectively. Compared to state-of-the-art HFC refrigerant with a volumetric cooling work of 12.87 MJm$^{-3}$ [4], our tension (48.63 MJm$^{-3}$) and bending (33.54 MJm$^{-3}$) results represent a 3.78 to 2.61 times improvement, respectively, suggesting potential improvements in system compactness when moving to eC cooling.

The observed reduction in self-heating and –cooling during bending as compared to tension is a result of the mechanics of bending beams. A bending moment causes variable displacement along the wire, resulting in maximum tension on one side, maximum compression on the other, and a neutral axis between those surfaces where the strain is zero. As a result, the material closest to the neutral axis undergoes less stress-induced martensitic transformation and a reduced temperature response. Additionally, the temperature response during bending increased nearly linearly from 2% to 8%, due to the aforementioned strain gradient in the material and access to untransformed material even at high strains. Interestingly, the difference in endothermic and exothermic response during bending did not change significantly at these higher strains (Figure 3e), as was the case during uniaxial testing (Figure 3d). This suggests improved slip, mechanical dissipation, and fatigue response during bending, although a more systematic study is required to prove this hypothesis.

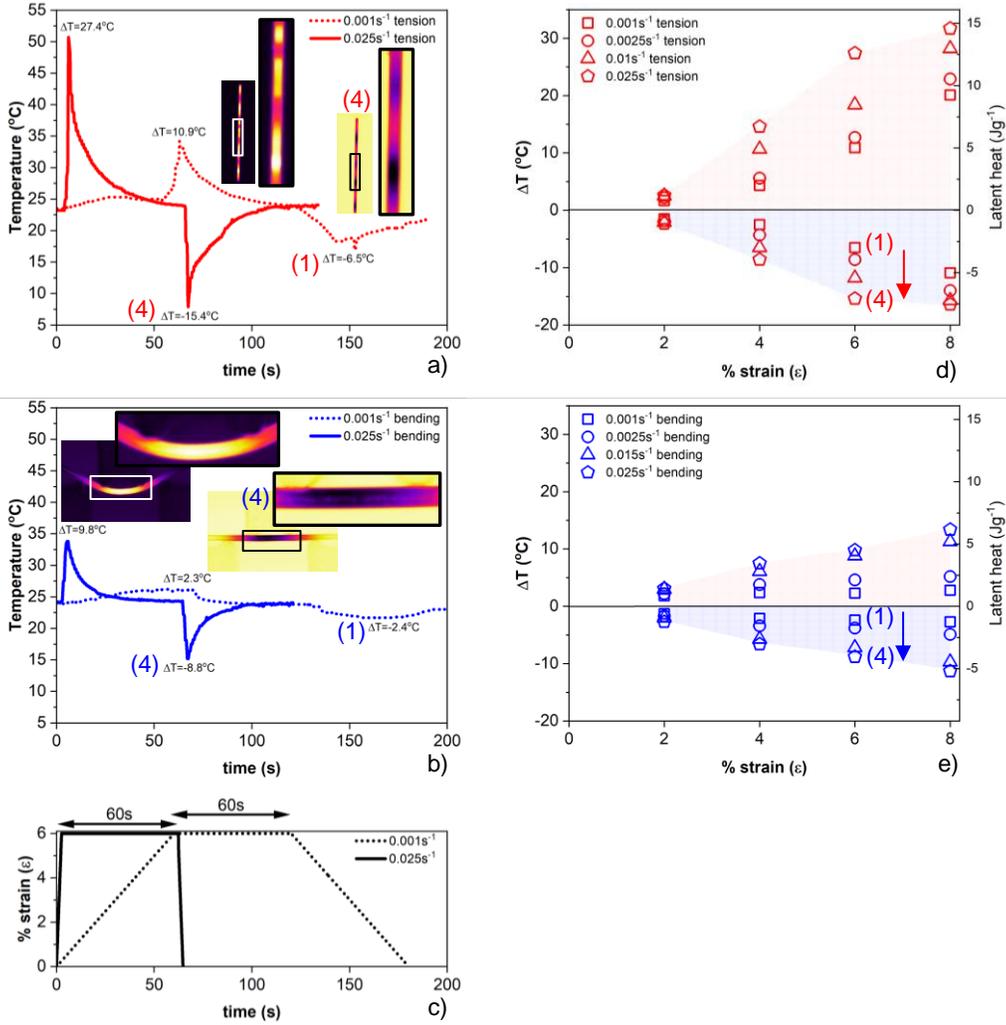

Figure 3: Thermal response during mechanical testing; a)-b) Temperature vs time plots for tension and bending tests at a strain of 6%. Inset IR images correspond with strain rate of 0.025s$^{-1}$; c) % strain vs time for minimum and maximum strain rates; d)-e) Temperature and calculated latent heat as a function of % strain and strain rate

The endothermic COP can be calculated by the quotient of the mass-specific latent heat (Figure 3d-e) and the mass-specific energy required to drive the hysteresis loop (Figure 2e-f):

$$\text{COP}_{cooling} = \frac{Q_{cool}}{Q_{hysteresis}} = \frac{mL_{endothermic}}{Fd} \qquad (2)$$

where $m$ is the mass of the sample undergoing phase transformation, $L_{endothermic}$ is the measured latent heat, $F$ is the applied force, and $d$ is the distance the force is applied. COP values for the tension and bending studies are shown in Figure 4 along with solid and dashed lines representing Carnot COP and 50% Carnot, respectively. Carnot COP ($T_c/(T_h-T_c)$) was calculated using the maximum ($T_h$) and minimum temperature ($T_c$) values for each condition in Figure 3d-e.

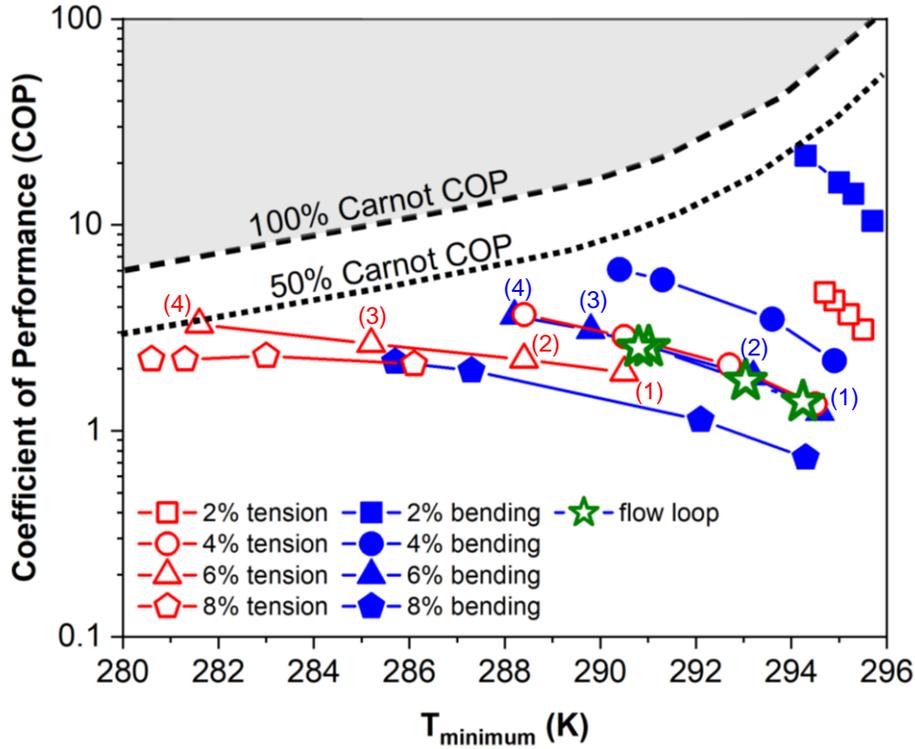

Figure 4: Calculated coefficient of performance for all tension and bending results, along with four experimental cases for the prototype

Numbered data points 1-4 on Figure 4 correspond to data points 1-4 on Figure 3d-e for a max strain of 6% and strain rates of 0.001 s$^{-1}$ and 0.025 s$^{-1}$. Generally, the observed ΔT (Figure 3) increased faster than increasing hysteresis work (Figure 2) as the strain rate increased, resulting in an improved COP. The bending and uniaxial results track well with Carnot COP, reducing as the temperature difference between endo- and exothermic temperatures increase. It's interesting to note that while uniaxial testing produced minimum temperature operation for a given maximum strain, bending provided equivalent or improved COP for a given cooling temperature; this is clear at a minimum temperature of 288.4K where 6% bending and 4% tension produced equivalent cooling temperature and near identical COP (3.6). Notably, the required force for this tension data point was 353.23 N (Figure 2b) while the required force during bending was only 60.85 N (Figure 2c). This represents a five-fold reduction in force for an equivalent COP and temperature lift by bending, rather than stretching, the SMA.

Motivated by the success of bending as a low-force and high-COP actuation mechanism, and the need for improved elastocaloric cooling designs, we developed a continuous elastocaloric cooling rotary-driven loop, as shown in Figure 5a. This setup consisted of a stepper motor, an 18mm-diameter copper tube to provide a maximum bending strain of ~6%, an Arduino controller, and a stepper motor driver. The un-stressed material, shown in Figure 5a, state [1], begins at room temperature in the Austenite phase. As the SMA is driven around the tube, it is forced to bend, thus inducing the exothermic austenite→martensite transformation (state [2]). Next, the released latent heat is dissipated to the copper tube and environment, thus cooling the stressed martensite material (state [3]). As the NiTi wire leaves the copper tube and un-bends, the reverse endothermic martensite→austenite transformation occurs and the alloy cools (state [4]). Here, the corresponding phase transformation may be used to absorb energy from the environment, or a discrete heat source, returning the un-stressed material to ambient temperature (state [1]). In this way, a continuous elastocaloric cooling loop resembling a reverse Brayton cycle is achieved. During testing, the feed rate (f) of the stepper motor (cm-s$^{-1}$) was adjusted to provide equivalent strain rates between 0.001 and 1.0 s$^{-1}$ to match conditions tested in the uniaxial and four point bending tests and extend operation to higher strain rate and higher power; please refer to the Supplementary Material for a more detailed description of the calibration procedure required to relate feed rate to motor speed.

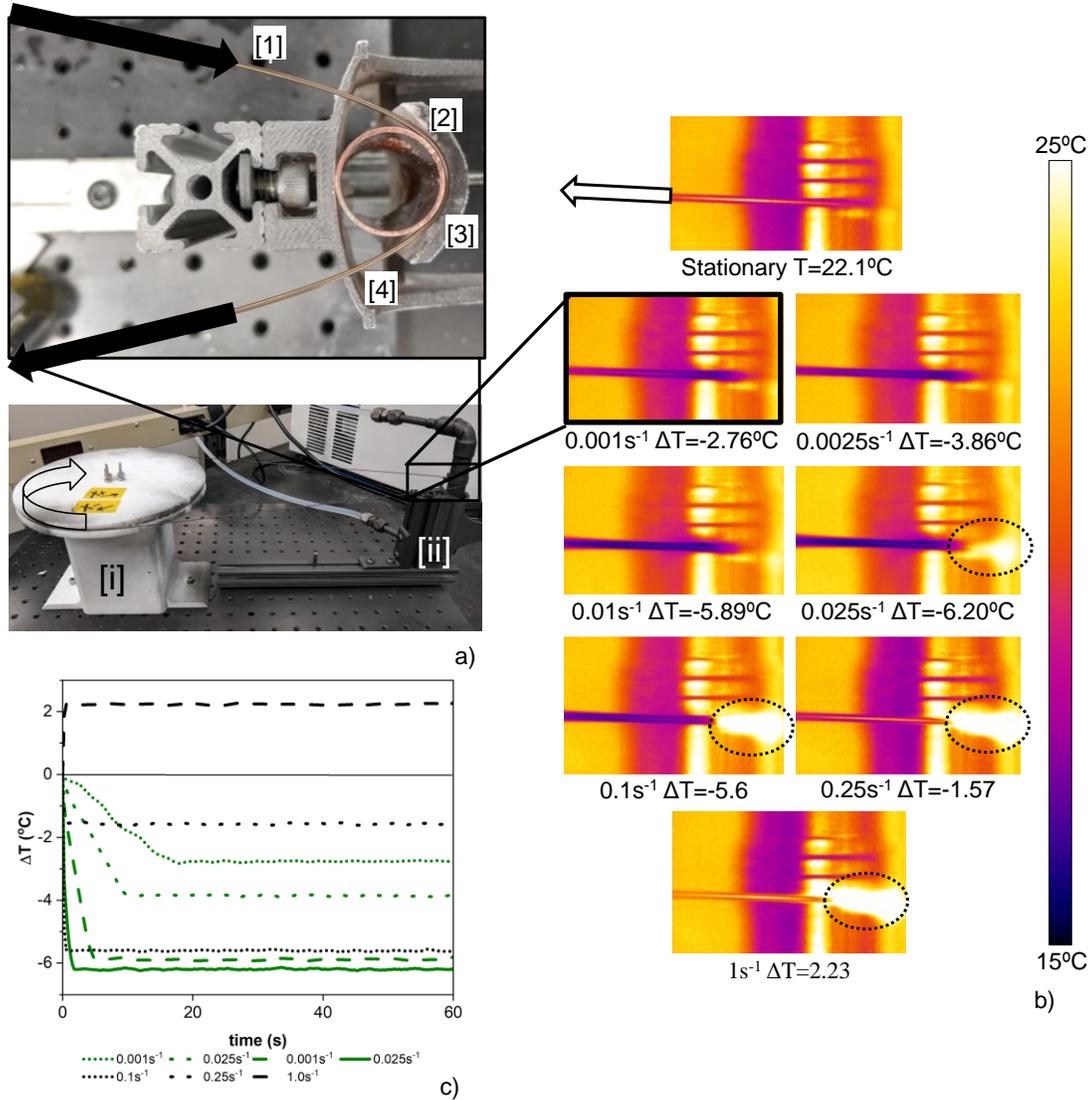

Figure 5: a) photograph and description of the proposed eC bending-mode continuous cooling prototype, b) IR images at strain rates from 0.001 to 1s$^{-1}$, and c) temperature vs time results at various strain rates.

Figure 5b shows infrared images of state [4] after 60 s of operation for strain rates ranging from 0.001 to 1 s$^{-1}$. For rates of 0.001 to 0.025 s$^{-1}$, the observed wire temperature starts at room temperature and decreases at varying rates before finally settling at temperatures between -2.76 to -6.20 °C as seen in Figure 5c. The observed settling time is a result of variable feed rates to accommodate the desired strain rates, whereby the highest strain rate has the highest feed rate and lowest settling time. For low strain rates ($\leq$0.0025 s$^{-1}$), the prototype temperature-drop values were within a few tenths of a degree to the bending results in Figure 3e. However, at larger strain rates, the temperature drop began to deviate significantly from the pure bending results. Specifically, at a strain rate of 0.01 s$^{-1}$ and 0.025 s$^{-1}$ the prototype temperature change was -5.89 ºC and -6.20 ºC which are 1.31 ºC and 2.6 ºC less than the four-point bending results, respectively. With further increase in strain rate, 0.1 to 1 s$^{-1}$, the prototype temperature change increased to -5.6, -1.57, and 2.23 ˚C. The reported temperature deviation is a result of several effects: frictional heating between the wire and copper tube and inadequate exothermic heat dissipation. The dwell time in the prototype was feed rate-dependent and varied from 19 s for the smallest strain rate to 35 ms for the highest strain rate, as opposed to 60 s during uniaxial and bending. Therefore, the exothermic latent heat was not removed from the sample before unloading occurred, thus reducing the observed temperature drop. This effect, combined with frictional heating, is apparent in Figure 5b at strain rates of 0.025 to 1 s$^{-1}$ where the sample is warmer than ambient. Nonetheless, assuming equivalent energy dissipation to pure bending, COP values from 1.38 to 2.55 can be reported for the prototype system with strain rates ranging from 0.001 to 0.025 s$^{-1}$ as shown on Figure 4.

In practical refrigeration and air conditioning applications, systems may experience tens-of-millions of loading and unloading cycles over their lifetime [4]. Past studies exploring rotary-bending fatigue properties in medical grade NiTi wire [32] and research into reducing fatigue in specially-designed and fabricated alloys [33] [34] [35] have demonstrated up to 10$^7$ tension and bending cycles. Notably, Zheng et al. [23] studied the relationship between Lüders bands, as shown in our uniaxial tension IR image in Figure 3a, and reported much shorter fatigue life in samples with bands than samples without. Using NiTi polycrystalline strip (55.89 wt.% Ni) from Johnson Matthey, Inc., they demonstrated that Lüders bands are nucleated at zones containing micro- and nano-sized defects. With repeated cyclic uniaxial loading and unloading, band nucleation and annihilation generates more damage due to the strong relationship between the localized martensitic phase transformation and the plasticity/defects formation [36] [37] [38]. In effect, Lüders bands are an indicator of accelerated material microstructure degradation, and are undesirable. Because Lüders bands were not observed in our bending samples (Figure 3b and Figure 5b), this indicates more-uniform and –homogeneous macroscopic phase transformations and potential for improved fatigue life. Future studies should explore long-duration and high-cycle bending performance for different maximum strains and strain rates using NiTi and consider alternative materials and manufacturing approaches to reduce fatigue [35].

We hypothesize that the continuous bending approach may be implemented in a parallel serpentine fashion, run at higher feed rate, and/or the cross-sectional area of the SMA can be increased to increase power according to the equation:

$$P_{cooling} = n\pi r^2 f \rho L_{endothermic} \qquad (3)$$

where n is the number of parallel loops and $\pi r^2$ represents the cross-sectional area of the wire. For example, at a proposed strain rate of 1 s$^{-1}$ and 100 parallel loops using the commercially-available NiTi wire (5.20 Jg$^{-1}$ and 1 mm-diameter), a cooling power of 1500 W (roughly 5000 BTU) would be theoretically possible. Because the proposed cooling prototype produces discrete (and continuous) hot and cold zones, much like standard vapor compression architectures, it is anticipated that active regeneration [39], in addition to passive recuperative regeneration schemes [40] and cascade designs [41] popularized by the VC industry can be adapted for our purposes to increase temperature lift. Studies are underway to elucidate these points.

This study has described the physics that govern performance of axisymmetric flexural bending as an actuation mechanism for eC cooling. Remarkably, these systematic bending tests revealed a five-fold reduction in actuation force compared to more-traditional uniaxial tension while maintaining equivalent COP and temperature lift. Furthermore, the absence of Lüders bands and reduced mechanical dissipation suggests improved fatigue life. The proposed eC loop design provides a logical path towards embracing bending-mode actuation while addressing scalability in size, power, and temperature lift by adjusting feed rate, sample cross section, parallelization, and active/passive regeneration schemes.

**SUPPLEMENTARY INFORMATION**

See supplementary material for information on Heckmann Diagram and interactions with external stimuli, eC thermodynamic cycles, methods for the determination of strain and strain rate during continuous bending prototype testing, and a more-detailed discussion of future work and additional material considerations.

**DATA AVAILABILITY**

The data that supports these findings are available from the corresponding author upon reasonable request.